\documentclass[12pt,a4paper]{article}%
\usepackage{amsmath}
\usepackage{amsfonts}
\usepackage{amssymb}
\usepackage{graphicx}%
\begin{document}

\begin{center}
\vspace{2cm}

{\Large \textbf{\textit{Singular inverse square potential in arbitrary
dimensions with a minimal length: Application to the motion of a dipole in a
cosmic string background}}} \vspace{1.3cm}

\textbf{Djamil Bouaziz\footnotetext{E-mail: djamilbouaziz@mail.univ-jijel.dz}%
}$^{1,2}$\textbf{ \& Michel Bawin}$^{2}$
\end{center}

\vspace{0.2cm}

\begin{center}
$^{1}$\textit{Universit\'{e} de Li\`{e}ge, Institut de Physique B5, Sart
Tilman 4000 Li\`{e}ge 1, Belgium.}

$^{2}$\textit{University of Jijel, Laboratory of Theoretical Physics, 18000,
Algeria.}
\end{center}

\vspace{0.5cm}

\hrule\vspace{0.5cm} {\normalsize \textbf{{\large {Abstract}}}} \vspace{0.5cm}

We solve analytically the Schr\"{o}dinger equation for the $N$-dimensional
inverse square potential in quantum mechanics with a minimal length in terms
of Heun's functions. We apply our results to the problem of a dipole in a
cosmic string background. We find that a bound state exists only if the angle
between the dipole moment and the string is larger than $\pi/4$. We compare
our results with recent conflicting conclusions in the literature. The minimal
length may be interpreted as a radius of the cosmic string.

\vspace{0.5cm} \hrule \vspace{2cm}

\section{Introduction}

We recently studied \cite{djamil} the attractive inverse square potential in
three dimensional quantum mechanics with a generalized uncertainty relation
implying the existence of a nonzero minimal uncertainty in position
measurement (minimal length) \cite{k1}. This study showed that this potential
remained regular in this framework; the elementary length, plays the role of a
regulator cutoff at short distances, and may be interpreted as an intrinsic
dimension of the system under study.

In this paper, we generalize the aforementioned work to $N$ dimensions and
arbitrary orbital momentum quantum number, and apply it to the problem of the
dipole dynamics in the background of a cosmic string, where the interaction is
known to be described by a two dimensional $1/R^{2}$ potential \cite{grats}.
Cosmic strings are very interesting one dimensional topological defects of
space-time \cite{davis}. Other types of defects are; point defects
(monopoles), planar defects ( domain walls) and textures. Such defects are
hypothesized to form in the phase transition of the early universe due to the
process of spontaneous symmetry breaking and some of them could have survived
to much later time, perhaps even to the present day \cite{davis,kibble,livre}.

The quantum dynamics of a point dipole in nonrelativistic quantum mechanics in
a cosmic string background has been considered by several authors (see, for
instance, Refs. \cite{lima1,lima2,grats,pulak}). In Ref. \cite{pulak}, the
author was interested in the case where the angle $\theta$ between the dipole
moment and the cosmic string is such that $\theta\leq\pi/4$, in which case the
potential is repulsive. The author used the method of self-adjoint extensions
\cite{metz} and found that the cosmic string can bind the dipole if the
potential is weakly repulsive, i.e., if one has $2M\delta/\hbar^{2}<1$, where
$M$ is the particle mass and $\delta$ is the strength of the potential. The
author claims that this result is an example of the classical scale symmetry
breaking of the system due to a "quantum anomaly". Note that from a
mathematical view point, a bound state may exist in a weakly repulsive
$1/R^{2}$ potential because the corresponding Hamiltonian has square
integrable solutions \cite{metz}. Given the counterintuitive feature of this
result (i.e., the existence of a bound state in a repulsive potential), it is
interesting to study the existence of bound states of that system in quantum
mechanics with a minimal length and to examine whether a cosmic string keeps
binding the dipole when the potential is repulsive.

The idea of modifying the standard Heisenberg uncertainty relation in such a
way that it includes a minimal length has first been proposed in the context
of quantum gravity and string theory \cite{amati}. It is assumed that this
elementary length should be on the scale of the Planck length of $10^{-35}$m,
below which the resolution of distances is impossible. The formalism based on
this modified uncertainty relation, together with the concepts it implies has
been discussed extensively by Kempf and his collaborators \cite{k1,k2}.
Various topics were studied over the last ten years within this formalism: the
hydrogen atom problem \cite{brau}, the harmonic oscillator potential
\cite{chang}, the Casimir effect \cite{nouicer}, the Dirac oscillator
\cite{kh} and the problem of a charged particle of spin one-half moving in a
constant magnetic field \cite{kh3}. The modifications of the gyromagnetic
moment of electrons and muons due to the minimal length have been discussed in
Ref \cite{sab1}. More recently several papers have been devoted to the study
of the black hole thermodynamics within the minimal length formalism
\cite{black}. For a review of different approaches of theories with a minimal
length scale and the relation between them, we refer the reader to Ref.
\cite{sab2}.

For the sake of completeness, let us mention that the $1/R^{2}$ interaction
that we study here in two spatial dimensions occurs in many problems of great
physical interest. Indeed, this potential appears in the study of electron
capture by polar molecules with static dipole moments \cite{bawin2,baw and c}.
The problem of atoms interacting with a charged wire is known to provide an
experimental realization of an attractive $1/R^{2}$ potential
\cite{johane,bawin1}. The Efimov effect in three-body systems \cite{efimov}
arises from the existence of a long range effective $1/R^{2}$\ interaction,
where $R$ is built from the relative distances between the three particles.
Finally, in black hole physics, the inverse square type interaction occurs
naturally in the analysis of the near-horizon properties of black holes, the
Bekenstein-Hawking entropy and black holes decay \cite{gupta}. Note finally
that the singular $1/R^{2}$ potential provides a simple example of a
renormalization group limit cycle in nonrelativistic quantum mechanics
\ \cite{beane}. Let us mention that the condition of square integrability of
the Schr\"{o}dinger wave function for a singular $1/R^{2}$ potential does not
lead to an orthogonal set of eigenfunctions with their corresponding
eigenvalues \cite{case,perelo,landau}. This is due to the fact that the
Hamiltonian operator is not essentially self-adjoint \cite{metz}, so that one
must define self-adjoint extensions of the Hamiltonian or equivalently require
orthogonality of the wave functions \cite{case}. The other technique used to
deal with this potential is the standard regularization by a cutoff at short
distances \cite{camblong}.

Our paper is organized as follows. In Sec. II, we study the $1/R^{2}$
potential in $N$ dimensional ($ND$) quantum mechanics with a minimal length,
using the momentum representation. In Sec. III, we study the problem of a
dipole in a cosmic string background. Our main result is that a bound state
exists only if the angle between the dipole and the cosmic string is larger
than $\pi/4$; the minimal length may be associated with the size of the cosmic
string. Some concluding remarks are reported in the last section.

\section{$N$ dimensional $1/R^{2}$ potential in quantum mechanics with a
generalized uncertainty relation}

In Ref. \cite{djamil} we have solved the $s$-wave Schr\"{o}dinger equation for
the three dimensional $1/R^{2}$ potential in quantum mechanics when the
position and momentum operators satisfy the following modified commutation relations:%

\begin{align}
\lbrack\widehat{X}_{i},\widehat{P}_{j}]  &  =i\hbar\lbrack(1+\beta\widehat
{P}^{2})\delta_{ij}+\beta^{^{\prime}}\widehat{P}_{i}\widehat{P}_{j}],\text{
\ \ }(\beta,\beta^{^{\prime}})>0.\nonumber\\
\lbrack\widehat{P}_{i},\widehat{P}_{j}]  &  =0,\label{1}\\
\lbrack\widehat{X}_{i},\widehat{X}_{j}]  &  =i\hbar\frac{2\beta-\beta
^{^{\prime}}+\beta(2\beta+\beta^{^{\prime}})\widehat{P}^{2}}{1+\beta
\widehat{P}^{2}}\left(  \widehat{P}_{i}\widehat{X}_{j}-\widehat{P}_{j}%
\widehat{X}_{i}\right)  .\nonumber
\end{align}

\bigskip These commutators imply the generalized uncertainty relation
\begin{equation}
\left(  \Delta X_{i}\right)  \left(  \Delta P_{i}\right)  \geq\frac{\hbar}%
{2}\left(  1+\beta\sum\limits_{j=1}^{N}[\left(  \Delta P_{j}\right)
^{2}+\left\langle \widehat{P}_{j}\right\rangle ^{2}]+\beta^{^{\prime}}[\left(
\Delta P_{i}\right)  ^{2}+\left\langle \widehat{P}_{i}\right\rangle
^{2}]\right)  . \label{2}%
\end{equation}
which leads to a lower bound of $\Delta X_{i}$, given by
\begin{equation}
\left(  \Delta X_{i}\right)  _{\min}=\hbar\sqrt{\left(  N\beta+\beta
^{^{\prime}}\right)  },\text{ \ \ }\forall i. \label{3}%
\end{equation}

Equation (\ref{2}) embodies the UV/IR mixing: when $\Delta P$ is large (UV),
$\Delta X$ is proportional to $\Delta P$ and, therefore, is also large (IR).
This phenomenon is said to be necessary to understand the cosmological
constant problem or the observable implications of short distance physics on
inflationary cosmology; it has appeared in several contexts: the AdS/CFT
correspondence, in noncommutative field theory and in quantum gravity in
asymptotically de Sitter space \cite{chang,sandore}. Another fundamental
consequence of the minimal length is the loss of localization in coordinates
space, so that, momentum space is more convenient in order to solve any
eigenvalue problem.

In the momentum representation, the following realization satisfies the above
commutation relations:
\begin{equation}
\widehat{X}_{i}=i\hbar\left(  (1+\beta p^{2})\dfrac{\partial}{\partial p_{i}%
}+\beta^{^{\prime}}p_{i}p_{j}\dfrac{\partial}{\partial p_{j}}+\gamma
p_{i}\right)  ,\text{ \ \ }\widehat{P}_{i}=p_{i}. \label{4}%
\end{equation}
The arbitrary constant $\gamma$ does not affect the observable quantities, its
choice determines the weight factor in the definition of the scalar product as
follow:
\begin{equation}
\left\langle \varphi\left\vert \psi\right.  \right\rangle =\int\frac{d^{N}%
p}{\left[  1+\left(  \beta+\beta^{^{\prime}}\right)  p^{2}\right]  ^{1-\alpha
}}\varphi^{\ast}(p)\psi(p),\text{ \ }\alpha=\frac{\gamma-\beta^{^{\prime}%
}\left(  \frac{N-1}{2}\right)  }{\beta+\beta^{^{\prime}}}. \label{5}%
\end{equation}

In the following, we generalize the work aforementioned to arbitrary
dimensions $N$ and arbitrary orbital momentum quantum number $l$.

We proceed, as in Ref. \cite{djamil}, by writing the Schr\"{o}dinger equation,
for a particle of mass $M$ in the external potential $V(R)=\delta/R^{2}$, in
the form
\begin{equation}
(R^{2}P^{2}+2M\delta)\left.  \left\vert \psi\right.  \right\rangle
=2MER^{2}\left.  \left\vert \psi\right.  \right\rangle \text{.} \label{6}%
\end{equation}

\bigskip Because of the rotational symmetry of the Hamiltonian, we can assume
that the momentum space energy eigenfunctions can be factorized as
\cite{chang}:
\begin{equation}
\psi_{N}(\vec{p})=Y_{l_{(N-1)}...l_{2}l_{1}}(\Omega)\varphi_{N}(p)\text{.}
\label{7}%
\end{equation}

Using Eq. (\ref{4}) with $\gamma=0$, we obtain the following expression for
$R^{2}\equiv\sum\limits_{i=1}^{N}X_{i}X_{i}$ \cite{djamil,chang}:%
\begin{align}
R^{2}  &  =\left(  i\hbar\right)  ^{2}\left\{  \left[  1+\omega_{1}%
p^{2}\right]  ^{2}\frac{d^{2}}{dp^{2}}+\left[  1+\omega_{1}p^{2}\right]
\left[  (N_{+}\beta+2\beta^{^{\prime}})p+\frac{N_{-}}{p}\right]  \frac{d}%
{dp}\right. \nonumber\\
&  \left.  -\frac{L^{2}}{p^{2}}-2\beta L^{2}-\beta^{2}L^{2}p^{2}\right\}  ,
\label{8}%
\end{align}
where we have used the notations
\[
\omega_{1}=\beta+\beta^{^{\prime}}\text{, \ \ \ }N_{\pm}=N\pm1\text{,
\ \ \ }L^{2}=l(l+N-2)\text{.}%
\]

From Eqs. (\ref{6}), (\ref{7}) and (\ref{8}) the radial Schr\"{o}dinger
equation for the $\delta/R^{2}$\ potential in the presence of a minimal length
takes the form%
\[
\frac{d^{2}\varphi_{N}(p)}{dp^{2}}+\left\{  \frac{4p}{p^{2}-2ME}+\frac
{(N_{+}\beta+2\beta^{^{\prime}})p+\frac{N_{-}}{p}}{1+\omega_{1}p^{2}}\right\}
\frac{d\varphi_{N}(p)}{dp}+
\]%
\begin{align}
&  +\left\{  2p\left[  \left(  (N+2)\beta+3\beta^{^{\prime}}\right)
p+\frac{N}{p}\right]  +\frac{1}{1+\omega_{1}p^{2}}\left[  -\beta^{2}L^{2}%
p^{4}+2\beta L^{2}(M\beta E-1)p^{2}\right.  \right. \nonumber\\
&  \left.  +\left.  (4M\beta E-1)L^{2}-\frac{2M\delta}{\hbar^{2}}%
+\frac{2MEL^{2}}{p^{2}}\right]  \right\}  \frac{\varphi_{N}(p)}{\left(
1+\omega_{1}p^{2}\right)  (p^{2}-2ME)}\nonumber\\
&  =0\text{.} \label{9}%
\end{align}

In the case $L=0$, this equation reduces to Schr\"{o}dinger equation of Ref.
\cite{djamil}.

Introducing the dimensionless variable $z$, defined as
\begin{equation}
z=\frac{(\beta+\beta^{^{\prime}})p^{2}-1}{(\beta+\beta^{^{\prime}})p^{2}+1},
\label{10}%
\end{equation}
which varies from $-1$ to $+1$, and using the following notations:%
\begin{equation}
\text{\ }\omega_{4}=\frac{\beta}{\beta+\beta^{^{\prime}}},\text{ \ }%
\omega=-M\omega_{1}E,\text{ \ }\kappa=\frac{M\delta}{2\hbar^{2}}\text{,}
\label{11}%
\end{equation}
we obtain the differential equation
\begin{align}
&  (1-z^{2})\frac{d^{2}\varphi_{N}}{dz^{2}}+\left\{  \left(  \frac{N_{+}%
\beta+2\beta^{^{\prime}}}{2\omega_{1}}-\frac{3}{2}\right)  (1+z)+\frac{N}%
{2}(1-z)+\frac{4(1+z)}{(1+2\omega)+(1-2\omega)z}\right\}  \frac{d\varphi_{N}%
}{dz}\nonumber\\
&  +\left\{  \frac{1}{1-z}\left[  -\left(  (\omega_{4}\omega+\frac{1}{4}%
)L^{2}+\kappa\right)  (1-z)^{2}-\frac{\omega_{4}^{2}L^{2}}{4}(1+z)^{2}%
+N_{-}(1-z)\right.  \right. \nonumber\\
&  \left.  \left.  +\frac{N_{+}\beta+2\beta^{^{\prime}}}{\omega_{1}%
}(1+z)+2\right]  -\dfrac{\frac{\omega L^{2}}{2}(1-z)^{2}}{1+z}-\frac
{\omega_{4}L^{2}}{2}(\omega_{4}\omega+1)(1+z)\right\} \nonumber\\
&  \frac{\varphi_{N}}{(1+2\omega)+(1-2\omega)z}\nonumber\\
&  =0\text{.} \label{12}%
\end{align}

To show that this equation can be transformed in the form of a Heun
differential equation, as in the $3D$ case with $l=0$ \cite{djamil},\ we make
the following transformation:
\begin{equation}
\varphi_{N}(z)=(1-z)^{\lambda}(1+z)^{\lambda^{^{\prime}}}f(z)\text{,}
\label{13}%
\end{equation}
where $\lambda$ and $\lambda^{^{\prime}}$ are arbitrary constants. Then, the
equation for $f(z)$ is
\begin{align}
&  \frac{d^{2}f}{dz^{2}}+\left\{  \frac{\tfrac{N_{+}\beta+2\beta^{^{\prime}}%
}{2\omega_{1}}-\tfrac{3}{2}-2\lambda}{(1-z)}+\frac{2\lambda^{^{\prime}}%
+\frac{N}{2}}{(1+z)}+\frac{4}{(1-z)\left[  (1+2\omega)+(1-2\omega)z\right]
}\right\}  \frac{df}{dz}\nonumber\\
&  +\frac{1}{(1-z^{2})^{2}\left[  (1+2\omega)+(1-2\omega)z\right]  }\left\{
\left[  (1+2\omega)+(1-2\omega)z\right]  \left[  \lambda(\lambda
-1)(1+z)^{2}\right.  \right. \nonumber\\
&  -2\lambda\lambda^{\prime}(1-z^{2})+\lambda^{\prime}(\lambda^{\prime
}-1)(1-z)^{2}-\lambda(\tfrac{N_{+}\beta+2\beta^{\prime}}{2\omega_{1}}%
-\tfrac{3}{2})(1+z)^{2}-\frac{N}{2}\lambda(1-z^{2})\nonumber\\
&  \left.  +\lambda^{\prime}(\tfrac{N_{+}\beta+2\beta^{\prime}}{2\omega_{1}%
}-\tfrac{3}{2})(1-z^{2})+\frac{N}{2}\lambda^{\prime}(1-z)^{2}\right]
+(\tfrac{N_{+}\beta+2\beta^{\prime}}{\omega_{1}}-4\lambda)(1+z)^{2}\nonumber\\
&  +(N_{-}+4\lambda^{\prime})(1-z^{2})-((\omega_{4}\omega+\frac{1}{4}%
)L^{2}+\kappa)(1+z)(1-z)^{2}\nonumber\\
&  \left.  -\frac{\omega_{4}L^{2}}{2}(\omega_{4}\omega+1)(1+z)^{2}%
(1-z)-\frac{\omega_{4}^{2}L^{2}}{4}(1+z)^{3}-\frac{\omega L^{2}}{2}%
(1-z)^{3}+2(1+z)\right\}  f\nonumber\\
&  =0\text{.} \label{14}%
\end{align}
\ 

We choose $\lambda$ and $\lambda^{^{\prime}}$ by requiring that the
coefficient \ of $f(z)$ in Eq. (\ref{14}) vanishes for $z=\pm1$; this leads to
the two equations for $\lambda$ and $\lambda^{^{\prime}}$ as follow:%
\begin{equation}%
\begin{array}
[c]{c}%
\lambda^{2}-(\frac{3}{2}+\tfrac{N_{+}\beta+2\beta^{\prime}}{2\omega_{1}%
})\lambda+\frac{1}{2}+\tfrac{N_{+}\beta+2\beta^{\prime}}{2\omega_{1}}%
-\frac{\omega_{4}^{2}L^{2}}{4}=0,\\
\lambda^{^{\prime}2}+(\frac{N}{2}-1)\lambda^{^{\prime}}-\frac{L^{2}}{4}=0.
\end{array}
\label{15}%
\end{equation}
The values of $\lambda$ and $\lambda^{^{\prime}}$ satisfying this system are
\begin{equation}%
\begin{array}
[c]{c}%
\lambda_{\pm}=\frac{1}{4}(3+\tfrac{N_{+}\beta+2\beta^{\prime}}{\omega_{1}}%
\pm\Delta_{1}),\\
\lambda_{\pm}^{\prime}=\frac{1}{2}(1-\frac{N}{2}\pm\Delta_{2}),
\end{array}
\label{16}%
\end{equation}
where
\begin{equation}
\Delta_{1}=\sqrt{\left(  \tfrac{N\beta+\beta^{\prime}}{\omega_{1}}\right)
^{2}+4\omega_{4}^{2}L^{2}},\text{ \ \ \ }\Delta_{2}=\sqrt{(\frac{N}{2}%
-1)^{2}+L^{2}}. \label{17}%
\end{equation}

We select the set $(\lambda,\lambda^{\prime})=(\lambda_{-},\lambda_{+}%
^{\prime})$; so the transformation (\ref{13}) becomes
\begin{equation}
\varphi_{N}(z)=(1-z)^{\frac{1}{4}(3+\tfrac{N_{+}\beta+2\beta^{\prime}}%
{\omega_{1}}-\Delta_{1})}(1+z)^{\frac{1}{2}(1-\frac{N}{2}+\Delta_{2})}f(z).
\label{18}%
\end{equation}

By substituting $\lambda$ and $\lambda^{^{\prime}}$ with their values in Eq.
(\ref{14}), we obtain after some calculations
\begin{equation}
\frac{d^{2}f}{dz^{2}}+\left\{  \frac{1-\frac{\Delta_{1}}{2}}{z-1}%
+\frac{1+\Delta_{2}}{z+1}+\frac{2}{z-z_{0}}\right\}  \frac{df}{dz}+\left\{
\frac{\rho z+\sigma}{(z-1)(z+1)(z-z_{0})}\right\}  f=0, \label{19}%
\end{equation}
where
\begin{align}
z_{0}  &  =\frac{2\omega+1}{2\omega-1},\nonumber\\
\rho &  =\frac{10-N}{4}+\frac{N(N\beta+\beta^{\prime})}{8\omega_{1}}%
-\frac{3\Delta_{1}}{4}+\frac{3\Delta_{2}}{2}-\frac{\Delta_{1}\Delta_{2}}%
{4}\nonumber\\
&  +\frac{1}{1-2\omega}\left\{  \left[  \frac{\omega_{4}^{2}}{2}-\omega
_{4}\omega+\frac{\omega_{4}}{2}(\omega_{4}\omega+1)\right]  L^{2}%
-\kappa\right\}  ,\label{20}\\
\sigma &  =\frac{1}{1-2\omega}\left\{  \frac{1}{4}(1+2\omega)\left(
\frac{N(N\beta+\beta^{\prime})}{2\omega_{1}}+2-N\right)  +\frac{1}{4}%
(2\omega-3)\Delta_{1}\right. \nonumber\\
&  \left.  +\frac{1}{2}(6\omega-1)\Delta_{2}-\frac{1}{4}(1+2\omega)\Delta
_{1}\Delta_{2}+\left[  \frac{\omega_{4}^{2}}{2}+\omega_{4}\omega+\frac
{\omega_{4}}{2}(\omega_{4}\omega+1)\right]  L^{2}+\kappa\right\}
\text{.}\nonumber
\end{align}

\bigskip Equation (\ref{19}) is a linear homogeneous second-order differential
equation with four singularities $z=-1,1,z_{0},\infty$, all regular. So, Eq.
(\ref{19}) belongs to the class of Fuchsian equations, and can be transformed
into the canonical form of Heun's equation, having regular singularities at
$z=0,1,\xi_{0},\infty$ \cite{snow,ronveau}. The simple change of variable
\[
\xi=\frac{z+1}{2}%
\]
leads to the following canonical form of Heun's equation:%
\begin{equation}
\frac{d^{2}f(\xi)}{d\xi^{2}}+\left(  \frac{c}{\xi}+\frac{e}{\xi-1}+\frac
{d}{\xi-\xi_{0}}\right)  \frac{df(\xi)}{d\xi}+\left(  \frac{ab\xi+q}{\xi
(\xi-1)(\xi-\xi_{0})}\right)  f(\xi)=0, \label{21}%
\end{equation}
with the parameters
\begin{align}
a  &  =\frac{3}{2}-\frac{\Delta_{1}}{4}+\frac{\Delta_{2}}{2}-\frac
{\widetilde{\nu}}{2},\text{ \ \ \ \ }\nonumber\\
b  &  =\frac{3}{2}-\frac{\Delta_{1}}{4}+\frac{\Delta_{2}}{2}+\frac
{\widetilde{\nu}}{2},\text{ \ \ \ }c=1+\Delta_{2},\text{ \ \ \ \ }d=2,\text{
\ \ \ \ }e=1-\frac{\Delta_{1}}{2},\text{ \ \ \ }\xi_{0}=\frac{2\omega}%
{2\omega-1},\nonumber\\
q  &  =-\frac{1}{1-2\omega}\left\{  1+(\frac{N}{4}-3)\omega-\frac{N(N-1)}%
{4}\omega_{4}\omega+\frac{\omega\Delta_{1}}{2}+(1-3\omega)\Delta_{2}%
+\frac{\omega\Delta_{1}\Delta_{2}}{2}-\omega_{4}\omega L^{2}-\kappa\right\}
,\nonumber\\
\widetilde{\nu}  &  =\sqrt{\left(  \frac{N-1}{2}\right)  ^{2}(\omega
_{4}-1)^{2}+\frac{1}{1-2\omega}\left\{  \left[  (1-2\omega)(1-2\omega
_{4})-\omega_{4}^{2}(4\omega+1)\right]  L^{2}+4\kappa\right\}  }\text{,}
\label{22}%
\end{align}
which are linked by the Fuchsian condition
\begin{equation}
a+b+1=c+d+e. \label{23}%
\end{equation}

In the neighborhood of $\xi=0$, the two linearly independent solutions of Eq.
(\ref{21}) are \cite{snow}
\begin{equation}
f_{1}(\xi)=H(\xi_{0},q,a,b,c,d;\xi), \label{24}%
\end{equation}%
\begin{equation}
f_{2}(\xi)=\xi^{1-c}H(\xi_{0},q^{\prime},1+a-c,1+b-c,2-c,d;\xi), \label{25}%
\end{equation}
where
\[
q^{\prime}=q-(1-c)\left[  d+\xi_{0}(1+a+b-c-d)\right]  .
\]
$H(\xi_{0},q,a,b,c,d;\xi)$ is the Heun function defined by the series
\begin{equation}
H(\xi_{0},q,a,b,c,d;\xi)=1-\frac{q}{c\xi_{0}}\xi+\sum\limits_{n=2}^{\infty
}C_{n}\xi^{n}, \label{26}%
\end{equation}
where the coefficients $C_{n}$\ are determined by the difference equation:%
\[
(n+2)(n+1+c)\xi_{0}C_{n+2}=\left\{  (n+1)^{2}(\xi_{0}+1)+(n+1)\left[
c+d-1\right.  \right.
\]%
\begin{equation}
\left.  +\left.  (a+b-d)\xi_{0}\right]  -q\right\}  C_{n+1}-(n+a)(n+b)C_{n},
\label{27}%
\end{equation}
with the initial conditions
\[
C_{0}=1,\ \text{\ }C_{1}=\frac{-q}{c\xi_{0}},\text{ and }C_{n}=0,\text{ if
}n<0.
\]

Now, we can write the solution of the deformed Schr\"{o}dinger equation
(\ref{9}) for the $N$ dimensional $1/R^{2}$ potential . Thus, by using Eq.
(\ref{18}) the solution $\psi(\xi)$, which is regular (finite) in the
neighborhood of $\xi=0$, is given by
\begin{equation}
\varphi_{N}(\xi)=A_{N}\xi^{\frac{1}{2}(1-\frac{N}{2}+\Delta_{2})}%
(1-\xi)^{\frac{1}{4}\left[  5+(N-1)\omega_{4}-\Delta_{1}\right]  }H(\xi
_{0},q,a,b,c,d;\xi), \label{28}%
\end{equation}
where $A_{N}$ is a normalization constant.

This formula generalizes that of Ref. \cite{djamil}, which can be recovered in
the special case: $N=3$ and $l=0$.

In the following section, we study in more detail the $2D$ case, by
considering a dipole in the presence of a cosmic string.

\section{Dipole dynamics in a cosmic string background}

Consider a particle of mass $M$, dipole moment $D$ moving in the background
field of a cosmic string. In non relativistic quantum theory the interaction
between the dipole and the cosmic string is described by the potential
\cite{grats,lima1,lima2,pulak}
\begin{equation}
V(R)=\frac{(1-\alpha^{2})D^{2}}{48\pi\alpha^{2}R^{2}}\cos2\theta, \label{35}%
\end{equation}
where $\theta$ is the angle between the string and the dipole moment and
$\alpha=1-4G\mu<1$ characterizes the cosmic string, with $\mu$ is the linear
mass density of the string and $G$ is the gravitational constant.

The potential (\ref{35}) is computed by considering the electromagnetic
self-energy of the dipole due to the non-flat geometry. The space-time metric
of the cosmic string background in cylindrical coordinate $(R,\phi,z)$ is
\cite{grats,lima1,lima2,pulak}
\begin{equation}
ds^{2}=dt^{2}-dz^{2}-dR^{2}-\alpha^{2}R^{2}d\phi^{2}\text{.} \label{36}%
\end{equation}

Because of the cylindrical symmetry of the space, the motion of the particle
along the $z$ direction is a free particle motion. By considering a cosmic
string of infinite length along the $z$ direction, one only has to discuss the
motion of the particle on the plane perpendicular to the $z$ direction.

The wave fuction of the dipole reads: $\psi_{2}(\vec{p})=e^{-im\phi}%
\varphi_{2}(p)$. The radial part, $\varphi_{2}(p)$, can be computed directly
from Eq. (\ref{28}) by setting $N=2$, and taking $\ 4\kappa=\frac
{M(1-\alpha^{2})D^{2}}{24\pi\alpha^{2}\hbar^{2}}\cos2\theta$. We obtain the
following expression:%
\begin{equation}
\varphi_{2}(\xi)=A\xi^{\frac{m}{2}}(1-\xi)^{\frac{1}{4}\left[  5+\omega
_{4}-\sqrt{(1+\omega_{4})^{2}+4\omega_{4}^{2}m^{2}}\right]  }H(\xi_{0}%
,q_{2},a_{2},b_{2},c_{2},d;\xi), \label{37}%
\end{equation}
with the parameters%
\begin{align}
a_{2}  &  =\frac{1}{4}\left(  6+2m-\sqrt{(1+\omega_{4})^{2}+4\omega_{4}%
^{2}m^{2}}\right)  -\frac{1}{2}\widetilde{\nu}_{2},\text{ \ \ \ \ }\nonumber\\
b_{2}  &  =\frac{1}{4}\left(  6+2m-\sqrt{(1+\omega_{4})^{2}+4\omega_{4}%
^{2}m^{2}}\right)  +\frac{1}{2}\widetilde{\nu}_{2},\text{ \ \ }\nonumber\\
\text{\ }c_{2}  &  =1+m,\text{ \ \ \ \ }d=2,\text{ \ \ \ \ }e_{2}=1-\frac
{1}{2}\sqrt{(1+\omega_{4})^{2}+4\omega_{4}^{2}m^{2}},\text{ \ \ \ }\xi
_{0}=\frac{2\omega}{2\omega-1},\nonumber\\
q_{2}  &  =\frac{-1}{1-2\omega}\left\{  1-\frac{\omega}{2}(5+\omega
_{4})+m(1-3\omega)-\omega\omega_{4}m^{2}+\frac{\omega}{2}(m+1)\sqrt
{(1+\omega_{4})^{2}+4\omega_{4}^{2}m^{2}}-\kappa\right\}  ,\nonumber\\
\widetilde{\nu}_{2}  &  =\sqrt{\frac{1}{4}(\omega_{4}-1)^{2}+\frac
{1}{1-2\omega}\left\{  4\kappa+\left[  (1-2\omega)(1-2\omega_{4})-\omega
_{4}^{2}(4\omega+1)\right]  m^{2}\right\}  }\text{.} \label{38}%
\end{align}

We now study the special case $m=0$ (ground state), and for convenience we
take $\beta^{\prime}=0$. In this case one has $e_{2}=0$ and $q_{2}=-a_{2}%
b_{2}$, so Heun's equation (\ref{21}) reduces to a hypergeometric equation
\cite{djamil} and the wave function becomes%
\begin{equation}
\varphi_{2}(\xi)=A(1-\xi)F(a_{2}^{\ast},b_{2}^{\ast},c_{2}^{\ast};\frac{\xi
}{\xi_{0}})\text{,} \label{40}%
\end{equation}
where
\begin{align}
a_{2}^{\ast}  &  =1-\frac{\widetilde{\nu}_{2}^{\ast}}{2}\text{,
\ \ \ \ \ \ \ \ }b_{2}^{\ast}=1+\frac{\widetilde{\nu}_{2}^{\ast}}{2}%
\text{,}\nonumber\\
c_{2}^{\ast}  &  =1\text{, \ \ \ \ \ \ \ \ \ \ \ \ \ \ \ \ \ \ }\widetilde
{\nu}_{2}^{\ast}=\sqrt{\frac{4\kappa}{(1-2\omega)}}\text{,}\label{41}\\
\xi &  =\frac{(\beta+\beta^{^{\prime}})p^{2}}{(\beta+\beta^{^{\prime}}%
)p^{2}+1}\text{, }\ \text{\ }\xi_{0}=\frac{2\omega}{2\omega-1}\text{.}%
\nonumber
\end{align}

We are now ready to investigate the existence of bound states. Let us observe
that, since $\cos2\theta$ varies from $-1$ to $+1$, the parameter $\kappa$ of
the dipole in the cosmic string background can be positive or negative. The
author of Ref. \cite{pulak} considered the repulsive case, i.e., $\theta
\leq\frac{\pi}{4}$, and using the method of self-adjoint extensions
\cite{metz}, found that the system would have a bound state in a weakly
repulsive potential ($0\leq4\kappa\leq1$). This bound state would be a
consequence of a "quantum anomaly" \cite{metz}.

As discussed in detail in Ref. \cite{djamil} for the $3D$ case, the physical
eigenfunctions of the Hamiltonian must behave at large momenta as
$p^{2}\varphi_{2}(p)\underset{p\rightarrow\infty}{=}0$. This boundary
condition emerged naturally from the integral equation corresponding to the
differential equation. It determines the physical behavior of the wave
function in this asymptotic region.We get from Eq. (\ref{40}) the following
quantization condition:
\begin{equation}
F(a_{2}^{\ast},b_{2}^{\ast},c_{2}^{\ast};\frac{2\omega-1}{2\omega})=0
\label{42}%
\end{equation}

In order to examine the existence of bound states for the dipole in a cosmic
string background, we have plotted the hypergeometric function in Eq.
(\ref{42}) as a function of $\omega=-M\beta E$ for fixed $4\kappa=$
$\frac{M(1-\alpha^{2})D^{2}}{24\pi\alpha^{2}\hbar^{2}}\cos2\theta$. The energy
eigenvalues $\omega_{n}$ are determined by the zeros of the function
$F(a_{2}^{\ast},b_{2}^{\ast},c_{2}^{\ast};\frac{2\omega-1}{2\omega})$ of Eq.
(\ref{42}). Figure 1 shows that there are no bound states for a weakly
repulsive potential. The parameters are taken identical to that used in Ref.
\cite{pulak}, namely $\theta=\pi/12$, $D=1$, $\alpha=0.2$ for the dashed curve
and $\theta=\pi/8$, $D=1.6$, $\alpha=0.2$ for the solid curve. In Ref.
\cite{pulak} one predicts one bound state depending on the value of the
self-adjoint extensions parameter ($\sum$). In quantum mechanics with a
minimal length, for any value of $\beta$ and $\beta^{\prime}$, there is no
bound state in the weakly repulsive potential. Figure 2 shows that in the case
$\theta=\pi/4$ (absence of interaction, i.e., $\kappa=0$) the cosmic string
cannot bind the dipole. In Ref. \cite{pulak}, on the other hand, one bound
state exists even if the strength of the potential is zero. \begin{figure}[h]
\begin{center}
\includegraphics[width=7cm,height=5cm]{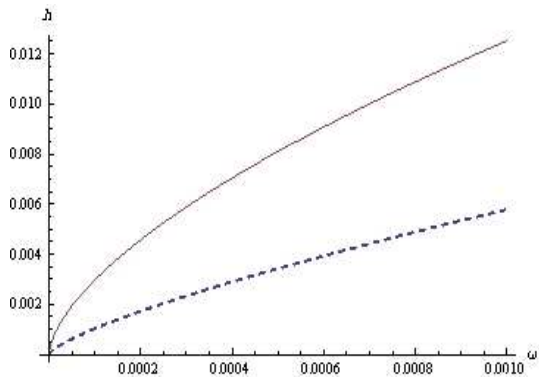}
\end{center}
\caption{$h\equiv F(a_{2}^{\ast},b_{2}^{\ast},c_{2}^{\ast};\frac{2\omega
-1}{2\omega})$ as a function of $\omega$, the dashed curve corresponds to
$4\kappa=0.2758$ and the solid curve corresponds to $4\kappa=0.5767.$ All
quantities $a_{2}^{\ast},b_{2}^{\ast},c_{2}^{\ast},$ $\omega,$ $\kappa$ are
dimensionless.}%
\label{Fig. 1}%
\end{figure}

Figures 3 and 4 show the appearance\ of bound states for $\theta>\pi/4$
(attractive potential). The energy of the ground state ($\omega_{0}$) is
finite; for $\kappa=-1/20$, $\omega_{0}=5.10^{-4}$ and for $\kappa=-3/2$,
$\omega_{0}=0.52$. As in Ref. \cite{lima2} for ordinary quantum mechanics,
there are many almost identical, excited states with $\omega\simeq0$
(accumulation point). In Fig. 4, we can see the energy of the first excited
state. \begin{figure}[h]
\begin{center}
\includegraphics[width=7cm,height=5cm]{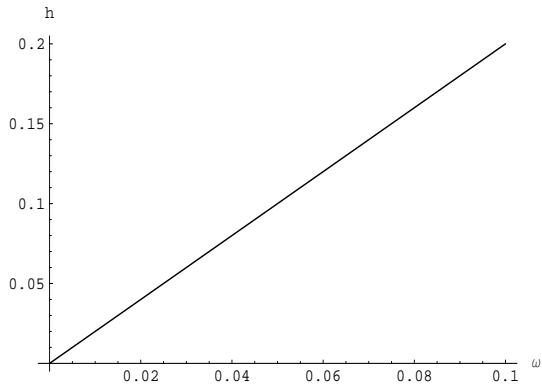}
\end{center}
\caption{$h\equiv F(a_{2}^{\ast},b_{2}^{\ast},c_{2}^{\ast};\frac{2\omega
-1}{2\omega})$ as a function of $\omega$, for $\kappa=0.$ All quantities
$a_{2}^{\ast},b_{2}^{\ast},c_{2}^{\ast},$ $\omega,$ $\kappa$ are
dimensionless.}%
\label{Fig. 2}%
\end{figure}\begin{figure}[hh]
\begin{center}
\includegraphics[width=7cm,height=5cm]{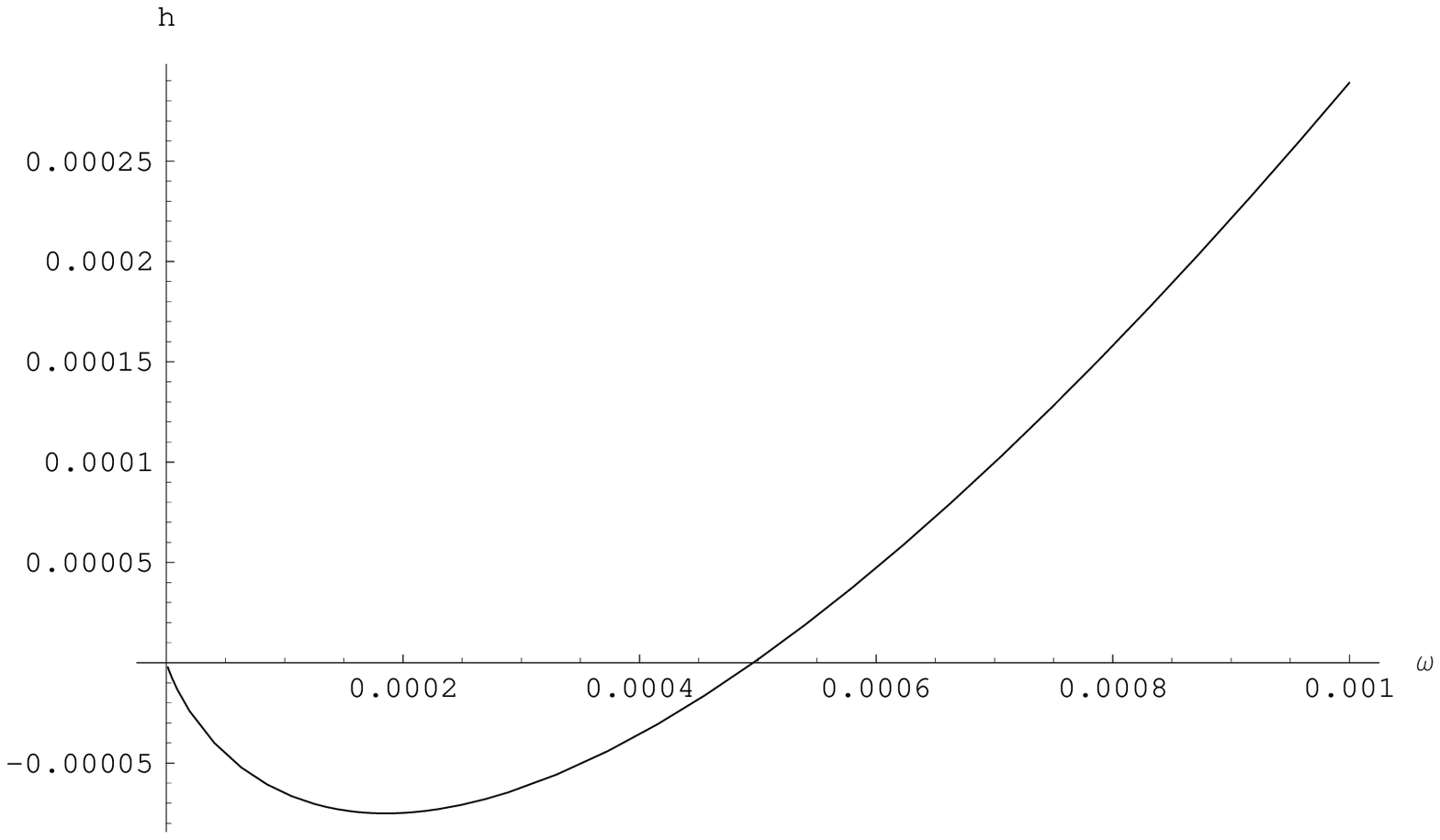}
\end{center}
\caption{$h\equiv F(a_{2}^{\ast},b_{2}^{\ast},c_{2}^{\ast};\frac{2\omega
-1}{2\omega})$ as a function of $\omega$, for $4\kappa=-1/5.$ All quantities
$a_{2}^{\ast},b_{2}^{\ast},c_{2}^{\ast},$ $\omega,$ $\kappa$ are
dimensionless.}%
\label{Fig. 3}%
\end{figure}\begin{figure}[hhh]
\begin{center}
\includegraphics[width=7cm,height=5cm]{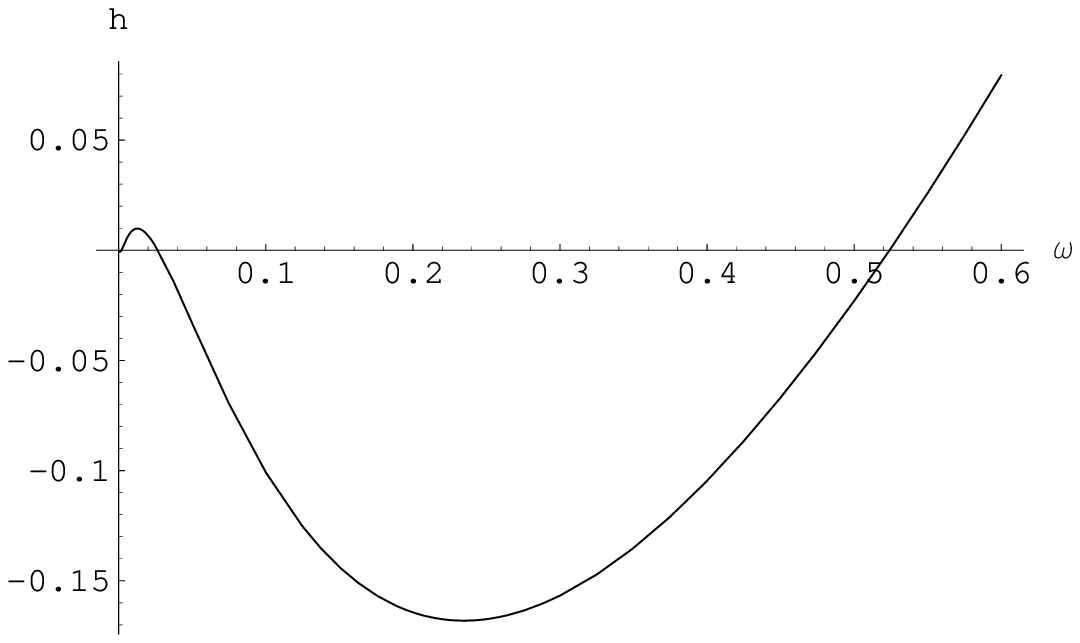}
\end{center}
\caption{$h\equiv F(a_{2}^{\ast},b_{2}^{\ast},c_{2}^{\ast};\frac{2\omega
-1}{2\omega})$ as a function of $\omega$, for $4\kappa=-6.$ All quantities
$a_{2}^{\ast},b_{2}^{\ast},c_{2}^{\ast},$ $\omega,$ $\kappa$ are
dimensionless.}%
\label{Fig. 4}%
\end{figure}

\vspace{2cm} \textbf{\bigskip}For the sake of completeness, we consider now
the case where $\theta>\frac{\pi}{4}$ ($\nu=$ $\sqrt{4\kappa}$ is now
imaginary) and a sufficiently small deformation parameter $\beta$ such that
$\omega=-\beta ME\ll1$. The quantization condition (\ref{42}) yields the
following expression for the energy levels \cite{djamil}:%
\begin{align}
E_{n}  &  =\frac{-\hbar^{2}}{M(\hbar\sqrt{2\beta})^{2}}\exp\left\{  \frac
{2}{\nu_{2}}\left[  \varphi-(n+\frac{1}{2})\pi\right]  \right\}  ,\label{43}\\
\nu_{2}  &  =\sqrt{-4\kappa},\text{ \ \ }\varphi=\arg\left[  \frac{\Gamma
(i\nu_{2})}{\Gamma(1+\frac{i\nu_{2}}{2})\Gamma(\frac{i\nu_{2}}{2})}\right]
\nonumber
\end{align}
as one has%
\[
\left\vert E_{n}\right\vert \ll\frac{1}{M\beta}\text{
\ \ \ \ \ \ \ \ \ \ \ \ \ \ \ \ \ }n=0,1,2,...\text{ .}%
\]

\bigskip This expression is similar to what was obtained in Ref. \cite{lima2},
in which a cutoff $a$ is introduced by hand to regularize the interaction at
short distances. In the above equation, the minimal length $(\Delta R)_{\min
}=\hbar\sqrt{2\beta}$ plays the same role as $a$. The author interpreted the
extra parameter $a$ as characterizing the radius of the string. We argue here
that this elementary length can be associated in this problem considered with
the finite size of the cosmic string.

\section{Summary and conclusion}

The problem of the $N$ dimensional singular inverse square potential has been
solved exactly for all values of the orbital momentum quantum number in the
framework of quantum mechanics with a minimal length. In the momentum
representation, the wave function is a Heun function for any dimension $N$,
and reduces to a hypergeometric function in some special cases. This result
generalize that of Ref. \cite{djamil}. As an application, we have considered a
dipole in a cosmic string background. This system is described by a
two-dimensional $1/R^{2}$ potential in non relativistic quantum theory, where
the coupling constant depends on the angle between the string and the dipole
moment ($\theta$). We have given the eigenfunctions of the Hamiltonian in the
presence of a minimal length, and the corresponding bound states equation. We
find that the cosmic string cannot bind the dipole for $\theta\leq\frac{\pi
}{4}$. This result is in contrast with that of Ref. \cite{pulak}. In the case
where $\theta>\frac{\pi}{4}$ we found that there exist many bound states and
the energy spectrum is bounded from below. We gave an expression for the
energy levels of bound states in the limit $\beta,$ $\beta^{\prime}\ll1$. Our
results agree with what was obtained in Ref. \cite{lima2}, where the same
problem is solved using the standard regularization technique by a cutoff
($a$) at short distances. The minimal length appears to be a natural regulator
and plays the same role as $a$. We argue with the author of Ref. \cite{lima2}
that this elementary length should be viewed as characterizing the finite
radius of the cosmic string.

\vspace{2cm} \textbf{Acknowledgments}: D. B acknowledges the Belgian Technical
Cooperation (BTC) and the Algerian ministry of Higher Education and Scientific
Research (MESRS) for their financial support.The work of M. B was supported by
the National Fund for Scientific Research (FNRS), Belgium.


\begin{thebibliography}{99}                                                                                               %


\bibitem {djamil}D. Bouaziz and M. Bawin, Phys. Rev. A \textbf{76}, 032112 (2007).

\bibitem {k1}A. Kempf, G. Mangano, and R. B. Mann, Phys. Rev. D \textbf{52},
1108 (1995).

\bibitem {grats}Y. Grats and A. Garc\'{\i}a, Class. Qantum. Grav. \textbf{13},
189 (1996).

\bibitem {davis}A.-C. Davis and T. W. B. Kibble, Contemp. Phys. \textbf{46},
313 (2005).

\bibitem {kibble}T. W. B. Kibble, J. Phys. A: Math. Gen. Vol. \textbf{9}, N.
8, 1387 (1976); T. W. B. Kibble, Phys. Rev. D, \textbf{26}, N. 2, 435 (1982).

\bibitem {livre}A. Vilenkin and E. P. S. Shellard, \textit{Cosmic Strings and
Other Topological Defects. }Cambridge University Press, Cambridge, U. K., 1994).

\bibitem {lima1}C. A. de Lima Ribeiro, C. Furtado, \ V. B. Bezera and F.
Moraes J. Phys. A: Math. Gen. \textbf{34}, 6119 (2001).

\bibitem {lima2}C. A. de Lima Ribeiro, C. Furtado, and F. Moraes, Mod. Phys.
Lett. A \textbf{20}, 1991 (2005).

\bibitem {pulak}P. R. Giri, Phys. Rev. A 76, 012114 (2007).

\bibitem {metz}K. Meetz, Nuovo Cimento \textbf{34}, 690 (1964).

\bibitem {amati}D. Amati, M. Ciafaloni, and G. Veneziano, Phys. Lett.
B\textbf{\ 216}, 41 (1989); M. Magiore, Phys. Lett. B\textbf{\ 319}, 83
(1993); L. J. Garay, Int. J. Mod. Phys. A\textbf{\ 10}, 145 (1995).

\bibitem {k2}A. Kempf, J. Math. Phys. \textbf{35}, 4483 (1994); H. Hinrichsen
and A. Kempf, J. Math. Phys. \textbf{37}, 2121 (1996); A. Kempf, J. Phys. A:
Math. Gen. \textbf{30}, 2093 (1997); A. Kempf, J. Math. Phys. \textbf{38},
1347 (1997); A. Kempf and G. Mangano, Phys. Rev. D \textbf{55} , 7909 (1997).

\bibitem {brau}F. Brau, J. Phys. A: Math. Gen. \textbf{32}, 7691 (1999); R.
Akhoury and Y.-P. Yao, Phys. Lett. B\textbf{\ 572,} 37 (2003); S. Benczik, L.
N. Chang, D. Minic, and T. Takeuchi, Phys. Rev. A \textbf{72}, 012104 (2005);
M. M. Stetsko and V. M. Tkachuk, Phys. Rev. A \textbf{74}, 012101 (2006); M.
M. Stetsko, Phys. Rev. A \textbf{74}, 062105 (2006).

\bibitem {chang}L. N. Chang, D. Minic, N. Okamura, and T. Takeuchi, Phys. Rev.
D \textbf{65}, 125027(2002)

\bibitem {nouicer}Kh. Nouicer, J. Phys. A: Math. Gen. \textbf{38}, 10027
(2005); U. Harbach and S. Hossenfelder, Phys. Lett. B \textbf{632}, 379 (2006).

\bibitem {kh}Kh. Nouicer, J. Phys. A: Math. Gen. \textbf{39}, 5125 (2006); C.
Quesne and V. M. Tkachuk, J. Phys. A : Math. Gen. \textbf{38}, 1747 (2005).

\bibitem {kh3}Kh. Nouicer, J. Math. Phys. \textbf{47}, 122102 (2006).

\bibitem {sab1}U. Harbach, S. Hossenfelder, M. Bleicher and H. Stoecker,
Proceedings of the Nuclear Physics Winter Meeting 2004, Bormio, Italy; e-print arXiv:hep-ph/0404205.

\bibitem {black}K. Nouicer, Phys. Lett. B 646, 63 (2007) ; Y-W. Kim and Y-J
Park, Phys. Lett. B V. 655, Issue. 3-4, 172 (2007); K. Nouicer, Class.
Quantum. Grav.25, 075010 (2008); Y-W. Kim and Y-J Park, Phys. Rev. D 77,
067501 (2008).

\bibitem {sab2}S. Hossenfelder, Class. Quant. Grav. \textbf{23}, 1815 (2006).

\bibitem {bawin2}M. Bawin, Phys. Rev. A\textbf{\ 70}, 022505 (2004).

\bibitem {baw and c}M. Bawin, S. A. Coon and B. R. Holstein, Int. Jour. Mod.
Phys A, Vol. \textbf{22}, N. 27, 4901 (2007), and references therein.

\bibitem {johane}J. Denschlag, G. Umshaus and J. Schmiedmayer, Phys. Rev.
Lett, \textbf{81}, 737 (1998).

\bibitem {bawin1}M. Bawin and S. Coon, Phys. Rev. A\textbf{\ 63}, 034701 (2001).

\bibitem {efimov}V. Efimov, Sov. J. Nucl. Phys. \textbf{12}. 589 (1971).

\bibitem {gupta}T. R. Govindarajan, V. Suneeta and S. Vaidya, Nucl. Phys. B
583, 291 (2000); D. Birmingham, K. S. Gupta and S. Sen, Phys. Lett. B.
\textbf{505}, 191 (2001); K. S. Gupta and S. Sen, Phys. Lett. B. \textbf{526},
121 (2002); K. S. Gupta and S. Sen, Phys. Lett. B. \textbf{574}, 93 (2003); K.
S. Gupta, Int. Jour. Mod. Phys A, Vol. \textbf{20}, N. 11, 2485 (2005).

\bibitem {beane}S. R. Beane, P. F. Bedaque et al, Phys. Rev. A\textbf{\ 64},
042103 (2001); M. Bawin and S. Coon, Phys. Rev. A\textbf{\ 67}, 042712 (2003);
E. Braaten and D. Phillips, Phys. Rev. A\textbf{\ 70}, 052111 (2004).

\bibitem {case}K. M. Case, Phys. Rev. \textbf{80}, 797 (1950).

\bibitem {perelo}A. M. Perelemov and V. S. Popov, Teor. Mat. Fiz. 4, 48 (1970)
[Theor. Math. Phys. \textbf{4}, 664 (1970)].

\bibitem {landau}G. H. Shortey, Phys. Rev, \textbf{38}, 120 (1931); F. L.
Scarf, Phys. Rev, \textbf{109}, 2170 (1958); W. M. Frank, D. J. Land and R. M.
Spector, Rev. Mod. Phys. \textbf{43}, 36 (1971); L. Landau and E. M. Lifshitz,
Quantum Mechanics, Vol. 3 (Pergamon Press, London, 1958) p. 118; P. M. Morse
and H. Feshbach, Methods of Theoretical Physics, Part II (McGraw-Hill, New
York, 1953), pp. 1665-1667.

\bibitem {camblong}K. S. Gupta and S. G. Rajeev, Phys. Rev. D\textbf{\ 48},
5940 (1993); H. E. Camblong, L. N. Epele, H. Fanchiotti, and C. A. Garc\'{\i},
Phys. Rev. Lett. \textbf{85}, 1590 (2000); S. A. Coon and B. R. Holstein, Am.
J. Phys. \textbf{70}, 513 (2002); H.-W. Hammer and B. G. Swingle, Anal. Phys,
\textbf{321}, 306 (2006).

\bibitem {abramo}Milton Abramowitz and Irene A. Stegum, Handbook of
Mathematical Functions With Formulas, Graphs, and Mathematical Tables; Fifth
Printing (U. S. Government Printing Office, Washington D. C., 1966), pp. 556-565.

\bibitem {sandore}S. Benczik et al, Phys. Rev. D \textbf{66}, 026003 (2002),
and references therein.

\bibitem {ronveau}A. Ronveaux, Heun's Differential Equations. Oxford, England
: Oxford University Press (1995).

\bibitem {snow}C. Snow, Hypergeometric and Legendre Functions With
Applications to Integral Equations of Potential Theory, National Bureau of
Standards Applied Mathematics Series (U. S. Government Printing Office,
Washington D. C. 1952), Vol. 19, pp. 87-101.
\end{thebibliography}
\end{document}